\documentclass[a4paper,12pt]{article}
\usepackage{color}
\usepackage{graphicx, color,epstopdf}
\usepackage{amsmath}
\usepackage{float}
\usepackage{tikz}
\usepackage[utf8]{inputenc}
\usepackage{hyperref}
\usepackage{authblk}
\usepackage{caption}

\usepackage{titlesec}
\usepackage{lipsum}
\usepackage[top=90pt,bottom=90pt,left=80pt,right=78pt]{geometry}

\usetikzlibrary{arrows,matrix,positioning}
\numberwithin{equation}{section}
\titleformat{\section}
{\normalfont\large\bfseries}{\thesection}{1em}{}

\title{\bf \large Uniqueness from gauge invariance and the Adler zero}
\author{Laurentiu Rodina}
\affil{Department of Physics, Princeton University, Princeton, NJ 08540}
\begin{document}
\maketitle
\begin{abstract}
In this paper we provide detailed proofs for some of the uniqueness results presented in Ref. \cite{Nima}. We show that: (1) Yang-Mills and General Relativity tree-level amplitudes are completely determined by gauge invariance in $n-1$ particles, with minimal assumptions on the singularity structure; (2) scalar non-linear sigma model and Dirac-Born-Infeld tree-level amplitudes are fixed by imposing full locality and the Adler zero condition (vanishing in the single soft limit) on $n-1$ particles. We complete the proofs by showing uniqueness order by order in the single soft expansion for Yang-Mills and General Relativity, and the double soft expansion for NLSM and DBI. We further present evidence for a greater conjecture regarding Yang-Mills amplitudes, that a maximally constrained gauge invariance alone leads to both locality and unitarity, without any assumptions on the existence of singularities. In this case the solution is not unique, but a linear combination of amplitude numerators.
\end{abstract}
\setcounter{tocdepth}{1}
\newpage

\tableofcontents

\section{Introduction}

Recently, in Ref. \cite{Nima} it was conjectured that after only fixing the number and form of possible singularities, gauge invariance uniquely determines the Yang-Mills and gravity scattering amplitudes. It was also stated that the same is true for scalar theories like the non-linear sigma model (NLSM) and Dirac-Born-Infeld (DBI), when gauge invariance is replaced by vanishing in the single soft limit. Crucially, in all of these cases locality and unitarity are never assumed, and so arise automatically as a consequence of uniqueness. Here by locality we mean that poles correspond to propagators of cubic diagrams (for YM and GR), and quartic diagrams (for NLSM and DBI), while by unitarity we mean factorization on those poles. In some sense then we see the emergence of spacetime and local quantum interactions purely from gauge invariance. A similar result was presented in Ref. \cite{bcfwl}, that locality and vanishing under large BCFW shifts are also sufficient to completely fix the Yang-Mills amplitude. Beyond their conceptual implications, these uniqueness results have a very practical application: if a given expression can be verified to be gauge invariant and contain the correct singularity structure, it is now guaranteed to match the corresponding amplitude. This has many implications for a wide variety of recently developed formalisms, like BCFW recursion relations \cite{BCFW}, the BCJ duality \cite{bcj}, or the CHY scattering equations \cite{CHY1}-\cite{CHY3}, among others.  

The proof used in this article also demonstrates a new powerful application of soft limits, as well as novel derivations of the well-known leading theorems \cite{Weinberg}\cite{doublesoftstuff}. Leading and subleading soft theorems have already proven very useful in a number of very surprising ways. Originally, they showed that charge conservation or the equivalence principle can be derived from S-matrix arguments \cite{Weinberg}. More recently, the theorems were interpreted as consequences of new symmetries \cite{strom1,strom2}, with further implications for black-hole information \cite{BH}. They were also used for recursion relations for effective field theories \cite{congkao}. 

The goal of this paper is three-fold. First, we present the full details of the argument used in Ref. \cite{Nima} to prove the uniqueness claims for the tree-level amplitudes of Yang-Mills, gravity, NLSM, and DBI, when locality is assumed. Second, we extend the argument to prove the conjecture that uniqueness still holds without assuming locality. And third, we make an even larger conjecture, that gauge invariance alone, with no assumptions on the presence of any singularities, is sufficient to imply both locality and unitarity.\footnote{This is similar to the results of \cite{boel1,boel2}, where it was found that imposing full gauge invariance, while also allowing extra kinematical invariants as coefficients, leads to $(n-3)!$ independent solutions, forming the BCJ basis of Yang-Mills amplitudes.}

\subsection{Assumptions and results}
In all four theories, Yang-Mills, gravity, NLSM, and DBI, we start with an ansatz $B_n(p^k)$, based on our assumptions of the singularity structure and mass dimension. In general, this ansatz contains functions of the form:
\begin{align}\label{nl}
B_n(p^k)\equiv \sum_{i} \frac{ N_i(p^k)}{P_i}\, ,
\end{align}
where the numerators $N(p^k)$ are general polynomials with $k$ powers of momenta, and linear in some number of polarization vectors/tensors (for YM/GR). The denominators $P_i$ can be any polynomial of $p_i.p_j$ factors. If we assume locality, it means we must restrict each $P_i$ to be a product of simple poles, which can be associated to propagators of cubic (for YM and GR) or quartic (for NLSM and DBI) diagrams. That is, a local ansatz has a form:
\begin{align}\label{ll}
B_n(p^k)\equiv \sum_{\textrm{diags. } i} \frac{ N_i(p^k)}{\prod_{\alpha_i}P_{\alpha_i}^2}\, ,
\end{align}
with $\alpha_i$ corresponding to the channels of each diagram. As discussed in \cite{bcj}, it is always possible to put amplitudes in this cubic diagram form, by adding artificial propagators to the higher point vertex interactions. For YM and NLSM the diagrams are ordered, while for gravity and DBI they are not. We can relax locality by dropping the underlying diagram structure, allowing each term to have some number $s$ of any singularities $P_\mathcal{S}^2=(\sum_i p_i)^2$, with the $p_i$ consecutive for YM and NLSM:
\begin{align}\label{lll}
B_n(p^k)\equiv \sum_{ i} \frac{ N_i(p^k)}{P_{\mathcal{S}_1}^2\ldots P_{\mathcal{S}_s}^2 }\, ,
\end{align}

Then the claim is that for the smallest $s$ and $k$ which admit solutions, the ansatz (\ref{lll}) is uniquely fixed by gauge invariance/vanishing in the single soft limit in $n-1$ particles. Concretely, these smallest values for $s$ and $k$ are:
\begin{itemize}
\item Yang-Mills: $s=n-3$, $k=n-2$
\item Gravity: $s=n-3$, $k=2n-4$
\item NLSM: $s=n/2-2$, $k=n-2$
\item DBI: $s=n/2-2$, $k=2n-4$
\end{itemize}

In this article, we prove the following results:
\begin{enumerate}
\item Local singularities + minimal mass dimension + gauge invariance $\Rightarrow$ Locality + Unitarity
\item Locality + minimal mass dimension + Adler zero $\Rightarrow$ Unitarity
\end{enumerate}
The stronger version of claim number 2 for NLSM and DBI (that Local Singularities + Adler zero $\Rightarrow$ Locality + Unitarity) is less susceptible to the argument we use in this article, but a more direct approach was already presented in \cite{Nima}. We also prove a stronger result for Yang-Mills, by allowing non-local singularities $(\sum_i a_i p_i)^2$, with some mild restrictions. Further, we conjecture that completely ignoring the singularity structure, gauge invariance alone forces general polynomials to be linear combinations of amplitude numerators.

Surprisingly, imposing gauge invariance/vanishing in the soft limit for the $n^{\rm th}$ particle is not required, and is automatic once the other $n-1$ constraints have been imposed. Without loss of generality we can take particle 3 to be this $n^{\textrm{th}}$ particle, and we will always impose momentum conservation by expressing $p_3$ in terms of the other momenta. Tying the unneeded $n^{\rm th}$ constraint to momentum conservation ensures that we always avoid checks of the form $e_3\rightarrow p_3=-p_1-p_2-p_4-\ldots$, which would complicate the analysis.

To begin, the above statements can be easily tested explicitly for a small number of particles. For Yang-Mills, at four points, we can only have terms with one pole, either $(p_1+p_2)^2$ or $(p_1+p_4)^2$. Then the most general term we can write down is a linear combination of 60 terms, of the form
\begin{align}
M_4(p^2)=a_1\frac{e_1.e_2 \,e_3.e_4\, p_1.p_4}{p_1.p_2}+a_2\frac{e_1.p_2\, e_3.p_2\, e_2.e_4}{p_1.p_2}+\ldots\, .
\end{align}
Imposing gauge invariance in particles 1, 2, and 4 forces the coefficients $a_i$ to satisfy some linear equations with a unique solution, which turns out to be precisely the scattering amplitude of four gluons.

At five points, the most general non-local ansatz, where we only assume two cyclic singularities per term, contains some 7500 terms, and it can be checked that gauge invariance in four particles leads to the five point amplitude. It is indeed quite remarkable that gauge invariance is so constraining to produce a unique solution. Actually, it is even more remarkable that any solution exists at all! The amplitudes are the result of a striking conspiracy between the propagator structure and momentum conservation.

It is easy to make a gauge invariant in $n$ particles by taking different contractions of $\prod_{i=1}^n (e_i^{\mu_i} p_i^{\nu_i}-e_i^{\nu_i} p_i^{\mu_i})$, but this requires a mass dimension of $[n]$. No single diagram has enough momenta in the numerator to accommodate this product, so different diagrams must cancel each other. But less obviously, without momentum conservation, such contractions will always contain terms with at least $n$ factors of $e_i.p_j$. This is impossible to achieve even with several diagrams, since each diagram numerator can have at most $n-2$ such factors per term, while the denominators only contains $p_i.p_j$ factors. But with momentum conservation, together with a cubic propagator structure, it turns out that $n-2$ factors are sufficient, creating an object which satisfies more constraints than expected by simple counting. In fact, we will show that this structure is indeed very special. There is no non-trivial way of deforming or adding things to produce different solutions. Identical facts hold for the other theories as well. For example, the NLSM requires vanishing under $n-1$ particles, which naively would require $A\propto \mathcal{O}(p_i)$ for $n-1$ particles. Again, however, only at $k=n-2$, with momentum conservation and quartic diagrams we find an exception, which is the amplitude itself. The absence of solutions below this critical mass dimension is at the heart of the proof.

The basic strategy for the proof is the following. We start with an appropriate ansatz (local, non-local, etc), and we take single/double soft expansions. Then order by order we show that gauge invariance/the Adler zero condition uniquely fix the corresponding amplitudes. In this process we do not assume any form of the soft theorems, but we end up re-deriving the well known leading terms \cite{Weinberg,doublesoftstuff}. Our approach does not directly provide the subleading terms (which for gravity have a particularly nice form \cite{strom3}), but only proves their uniqueness.\footnote{See also \cite{delta} for a very illuminating discussion on fixing the subleading terms.} The whole proof rests on showing that after the first non-vanishing order is fixed, none of the higher orders can produce independent solutions. The reason for this is that the subleading orders must have a growing number of soft momenta in the numerator, leaving fewer momenta to satisfy the necessary requirements.

\subsection{Organization of the article}
In section \ref{gaugeinv}, we begin by exploring the notion of constrained gauge invariance. We find a very simple proof that functions with at most $k<n-2$ factors of momenta in the numerator can be gauge invariant in at most $k$ particles, while the same is true for tensors with $k\le n-2$.

In section \ref{unique1}, we first prove a weaker version of our statement for YM and GR, by assuming locality. In section \ref{unique2} the same argument is applied to the NLSM and DBI amplitudes, with gauge invariance replaced by the Adler zero condition.

In section \ref{locality}, relaxing our assumptions on the underlying cubic diagrams, we instead consider a more general singularity structure. We only keep the requirement of (local) singularities of the form $(\sum_i p_i)^2$, and recover the unique amplitude, as long as the number of such singularities per term is $s= n-3$. For fewer singularities there are no solutions, while for more the answer can always be factorized as $(\sum \textrm{poles})\times \textrm{(amplitude)}$. This proves the conjecture originally made in Ref \cite{Nima}.

Finally, in section \ref{l2}, we investigate the extent to which more general singularities can be used to fix the Yang-Mills amplitude. Completely non-local singularities of the form $(\sum_i a_i p_i)^2$, with some minor restrictions, are also shown to provide a unique solution. Trying to find a less arbitrary ansatz, we are lead to consider polynomials again, with no singularities at all. We conjecture that yet an even stronger statement can be made, namely that the smallest mass dimension polynomial that admits a solution is fixed to a linear combination of amplitude numerators, when gauge invariance in {\em all} $n$ particles is imposed. The usual argument can be used to provide leading order evidence for this fact.

\section{Constrained gauge invariance}\label{gaugeinv}

\subsection{Polynomials}
Let $B(k)$ be a polynomial linear in polarization vectors, with at most $k$ factors of dot products of the type $e_i.p_j$ in any given term. Let $g$ be the total number of gauge invariance requirements, and $\Delta=g-k$ be the ``excess" of gauge invariance requirements compared to the maximum number of $e.p$ factors. When appropriate, we will use the notation $\overline{e}_i$ to distinguish polarization vectors which are not used for gauge invariance. We wish to prove that, without momentum conservation, $B(k)$ can be gauge invariant in at most $k$ particles, ie. satisfy at most $\Delta=0$ constraints. Then we will prove that with momentum conservation the statement is still true, but only for $k<n-2$.

\subsubsection{No momentum conservation}
Having no momentum conservation implies gauge invariants in particle $i$ must be proportional to $G_i^{\mu\nu}=e_i^\mu p_i^\nu-e_i^\nu p_i^\nu$. Therefore the only way to obtain gauge invariants in $k$ particles is with linear combinations of different contractions of products $\prod_i G_i^{\mu_i\nu_i}$. We will show that such expressions always contain at least one term with $k$ factors of $e.p$.

By assumption there are always more $e$'s than $e.p$'s, so at least one of the polarization vectors needed for gauge invariance will be in a factor $e.e$ or $e.\overline{e}$. Consider first as an example the following term in a polynomial with $k=2$:
\begin{align}
\label{eq0}e_1.e_2\ e_3.p\ \overline{e}_4.p\ p.p\, .
\end{align}
We wish to show that such a term cannot be gauge invariant in three particles, say particles 1, 2 and 3. We start with a polarization vector sitting in a $e.e$ factor, for example $e_1$. To make a gauge invariant in particle 1, at least one of the $p's$ above must be a $p_1$, and a pair term must exist with $e_1$ and a $p_1$ switched. We can use either the $p$ in the $e_3.p$ (or $\overline{e}_4$) factor, or one in the $p.p$ factor, to make the gauge invariants:
\begin{align}
\label{eq1}
G_1&= (e_1.e_2)\ (e_3.p_1)\ \overline{e}_4.p\ p.p - (p_1.e_2)\ (e_3.e_1)\ \overline{e}_4.p\ p.p\, ,\\
\nonumber &\textrm{or}\\
\label{eq2}G_1'&=\ (e_1.e_2)\ e_3.p\ \overline{e}_4.p\ (p_1.p)- (p_1.e_2)\ e_3.p\ \overline{e}_4.p \ (e_1.p)\, .
\end{align}
However, the second option leads to a term with four $e.p$ factors, contradicting our claim that just two factors are sufficient. The first option is fine, and so we can only use $p$'s in $e.p$ factors for this mark-and-switch procedure.

Now consider the second term in $G_1$ above, and note that we ended up with another $e.e$ factor, namely $e_3.e_1$. Applying the same reasoning for gauge invariance in 3 forces us to fix the factor $\overline{e}_4.p$ to $\overline{e}_4.p_3$. Therefore the second piece of $G_1$ can form a gauge invariant in particle 3 in the pair:
\begin{align}
G_3=(p_1.e_2)\ (e_3.e_1)\ \overline{e}_4.p_3\ p.p-p_1.e_2\ p_3.e_1\ \overline{e}_4.e_3\ p.p\, .
\end{align}
Now we do not need gauge invariance in 4, so this chain $1\rightarrow 3\rightarrow \overline{4}$ is finished. Note we do not care about making the first piece of $G_1$ gauge invariant in particle 3, we are only interested in finding some minimal constraints. Instead, we go back to (\ref{eq0}) to check gauge invariance in the remaining particle 2. But the choices we made so far by imposing gauge invariance in 1 and 3 fixed both $e.p$ factors in this initial term to
\begin{align}
e_1.e_2\ e_3.p_1\ \overline{e}_4.p_3\ p.p\, ,
\end{align}
so now there is no way to make it gauge invariant in 2, as all the allowed $p$'s have been used up. Therefore the term (\ref{eq0}) is not compatible with gauge invariance in \{1,2,3\}.

The general strategy is the same. New chains always start in the original term from $e.\overline{e}$ or $e.e$ factors, which are aways present by assumption. Next, for each jump we fix a $p$ in an $e.p$ or $\overline{e}.p$ factor, which becomes unavailable for other gauge invariants. The chain ends when reaching an $\overline{e}.p$ factor, and a new chain is started from the original term, and so on. The process ends when all the chains have ended on $\overline{e}.p$ factors, which means gauge invariance is compatible with this counting argument, or when all the $e.p$'s have been marked and a chain is unable to continue, which means the term is ruled out.

In general, we said we need gauge invariance in $k+1=\#[\overline{e}.e]+2\#[e.e]+\#[e.p]$ particles, but have only $k=\#[\overline{e}.p]+\#[e.p]$ factors. This means the difference between how many chains must start and how many can end is: $\left(\#[\overline{e}.e]+2\#[e.e]\right)-\#[\overline{e}.p]=1$. Therefore there is always at least one chain which cannot end, so all possible starting terms are ruled out. Then there is no way to make a polynomial $B(k)$ gauge invariant in $k+1$ particles.

\subsubsection{With momentum conservation}
Now we consider the same type of polynomials from above, but on the support of a momentum conserving delta function, $B_n(k)\equiv B(k)\delta_n$, where $\delta_n\equiv \delta(\sum_i p_i)$. To impose momentum conservation explicitly, we can choose three particles (for example 2, 3, and 4), and use the following relations:
\begin{align}
p_3&=-\sum_{i\neq 3} p_i\, ,\\
e_3.p_4&=-\sum_{i\neq 3,4} e_3.p_i\, ,\\
p_2.p_4&=-\sum_{i,j\neq 3} p_i.p_j\equiv P_{24}\, .
\end{align}
This allows other ways of forming gauge invariance in particles 2, 3 and 4. For example, for particle 2 we can now have a new gauge invariant of the form
\begin{align}\label{alt}
G_2^\mu =e_2.p_4 p_2^\mu-P_{24} e_2^\mu\, .
\end{align}
Such an expression avoids our previous argument: now both particles 2 and 4 can share the same $e_2.p_4$ factor above. When checking gauge invariance in at most $n-3$ particles this is not a problem, as we can always impose momentum conservation in such a way as to avoid these three special particles. However, for more than $n-3$ particles, at least one particle has to be affected by momentum conservation.

The worst case that we will need to prove is that $B_n(n-3)$ cannot be gauge invariant in $n-2$ particles. Without loss of generality, since we can change how we impose momentum conservation, we can leave out 3 and 4, and assume that these $n-2$ particles are $\{1,2,5,6,\ldots,n\}$. Then $n-3$ of the particles can only form gauge invariants of the form $ e_i^\mu p_i^\nu-e_i^\nu p_i^\nu$, while particle 2 allows gauge invariants of the form (\ref{alt}). Since we only have $k=n-3$ factors of $e.p$, the first $n-3$ constraints already fix all such factors. However, we are not checking gauge invariance in particle 4, so none of the factors will be fixed to $e.p_4$, more specifically to the $e_2.p_4$ needed in eq. (\ref{alt}). Therefore there is still no room to form the gauge invariant for particle 2.

This proves that a polynomial $B_n(k)$ with $k<n-2$ can be gauge invariant in at most $k$ particles. It is easy to see that the last step of the argument fails for $k=n-2$. In that case the counting allows for gauge invariance in not just $n-1$ particles, but all the way to $n$ particles. This is of course what we should expect, since a polynomial $B_n(n-2)$ corresponds to the full amplitude numerator, and is gauge invariant in $n$ particles.

\subsection{Functions and tensors with singularities}
The previous results for polynomials can be extended to functions with poles, such as those we initially introduced in eq. (\ref{nl}):
\begin{align}
\label{nonlocal}
B_n(p^k)\equiv \sum_{i} \frac{ N_i(p^k)}{P_i}\, .
\end{align}
Because the $P_i$ are only functions of $p.p$ factors, a function with at most $k$ momenta in the numerators can be expressed in terms of a polynomial with at most $k$ factors of $e.p$:
\begin{align}
B_n(p^k)=\frac{B_n(k)}{\prod_i P_i}\, .
\end{align}
This implies that a function $B_n(p^k)$ cannot be gauge invariant in $k+1$ particles, if $k< n-2$. This statement can be generalized to tensors $B^{\mu\nu}_n(p^k)$. We can write out the components of such a tensor:
\begin{align}
\label{qq1}B_n^{\mu\nu}(p^k)&=\sum_{i, j}p_i^\mu p_j^\nu B_{ij}(p^{k-2})+\sum_{i,j} p_i^\mu e^\nu_j C_{ij}(p^{k-1})+\sum_{i,j} e_i^\mu p^\nu_j C'_{ij}(p^{k-1})+\sum_{i\neq j} e^\mu_i e^\nu_j D_{ij}(p^{k})\, ,
\end{align}
and determine what constraints each of the functions above must satisfy in order for $B^{\mu\nu}_n(p^k)$ to be gauge invariant in $k+1$ particles. We can treat each different type of function in order:
\begin{itemize}
\item $p_i^\mu p_j^\nu B_{ij}(p^{k-2})$: if we check gauge invariance in some particle $m$, with $m\neq i,j$, the prefactor remains unique, and none of the other terms in (\ref{qq1}) may cancel against $B_{ij}$. This implies $B_{ij}(p^{k-2})$ itself must be gauge invariant in at least $k+1-|\{i,j\}|=k-1$ particles so is ruled out, if $k-2< n-2$.

\item $e_i^\mu p_i^\nu C_{ii}(p^{k-1})$, or $\overline{e}_i^\mu p_j^\nu C_{ij}(p^{k-1})$: the same logic as before implies $C(p^{k-1})$ must be gauge invariant in $k$ particles, so is also ruled out if $k-1<n-2$.

\item $e_i^\mu p_j^\nu C_{ij}$: can only form a gauge invariant in $i$ together with a term $p_i^\mu p_j^\nu B_{ij}$, which was ruled out.

\item $e_i^\mu e_j^\nu D_{ij}(p^k)$, or $e_i^\mu \overline{e}_j^\nu D_{ij}(p^k)$: under $e_i\rightarrow p_i$, $D_{ij}$ can only form a gauge invariant with $C_{ij}$, which vanished, so this case is ruled out.
\item $\overline{e}_i^\mu \overline{e}_j^\nu D_{ij}(p^k)$: is only ruled out for $k<n-2$
\end{itemize}
To summarize, we obtain just three types of cases:
\begin{align}
&B_n(p^{k-2}), \textrm{ gauge invariant in $k-1$}\, ,\\
&C_n(p^{k-1}), \textrm{ gauge invariant in $k$}\, ,\\
\label{3rd}&D_n(p^{k}), \textrm{ gauge invariant in $k+1$}\,
\end{align}
all of which vanish for $k< n-2$. For $k=n-2$, the first two also vanish, but for the third we have an apparent contradiction, since we know that functions $D_n(p^{n-2})$ have sufficient momenta to satisfy gauge invariance in $n-1$ particles. However, when $k=n-2$, case (\ref{3rd}) does not exist. The only tensor $B_n^{\mu\nu}(p^{n-2})$ that will show up in the actual proof has $\overline{G}=\{e_3\}$, so it does not contain a component $\overline{e}_i\overline{e}_j D_{ij}$.

Therefore, the tensors we are interested in cannot be gauge invariant in $k+1$ particles for $k\le n-2$. In conclusion, so far we have shown that:
\begin{itemize}
\item functions $B_n(p^k)$ cannot satisfy $\Delta=1$ constraints for $k<n-2$
\item tensors $B^{\mu\nu}_n(p^k)$ cannot satisfy $\Delta=1$ constraints for $k\le n-2$
\end{itemize}
It turns out that these results can be generalized even further: we can take linear combinations of the above functions/tensors, with factors of $p_i.p_j$ as coefficients, and still the above statements hold. These results will be necessary for the following sections.

\section{Unitarity from locality and gauge invariance}\label{unique1}
In this section we will consider local functions, as in eq. (\ref{ll}):
\begin{align}
\label{local}
B_n(p^{k})= \sum_{\textrm{diags. } i} \frac{ N_i(p^{k})}{\prod_{\alpha_i}P^2_{\alpha_i}}\, .
\end{align}
In the above notation the actual gluon amplitude $A_n(p^{n-2})$ is a subset of $B_n(p^{n-2})$, with $G=\{1,2,4,\ldots,n\}$ and $\overline{G}=\{3\}$, so $g=n-1$ and $\Delta=1$. Now we wish to prove that $A_{n+1}$ is uniquely fixed by gauge invariance in $n$ particles, under the assumption that $A_n$ is fixed by gauge invariance in $n-1$ particles. Consider the most general $(n+1)$-point local function $M_{n+1}$, and let $p_{n+1}=z q$. The Taylor series expansion around $z=0$ is:
\begin{align}
\nonumber M_{n+1}\delta_{n+1}=&(z^{-1} M^{-1}_{n+1}+z^0 M_{n+1}^0+\ldots)(\delta_n+z q.\delta_n'+\ldots)\\
\nonumber \label{Taylor}=&z^{-1}M_{n+1}^{-1}\delta_n+z^0\left(M_{n+1}^{-1}q.\delta'_n+M^0_{n+1}\delta_n\right)+\ldots\\
=&z^{-1}\mathcal{M}_{n+1}^{-1}+z^0\mathcal{M}_{n+1}^0+\ldots\, .
\end{align}
First we must investigate the pole structure of a local function in this limit. There are two types of poles that can show up. First, there are $q$-poles which are singular. These correspond to diagrams of the type:
\begin{figure}[H]
\centering
\includegraphics[scale=0.5]{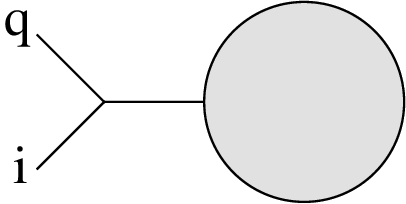}
\end{figure}
\noindent and can be written as:
\begin{align}
\frac{N}{q.p_i \mathcal{P}_n(q)}=D_{n+1}
\end{align}
with $i=1$ or $i=n$ for Yang-Mills, because of ordering. In the limit $q\rightarrow 0$, we can factor out the $q.p_i$ pole and incorporate the remaining propagator structure into the lower point local function:
\begin{align}\label{d1}
D_{n+1}\rightarrow \frac{1}{q.p_i} \frac{N}{\mathcal{P}_n(0)}=\frac{1}{q.p_i} B_n\, .
\end{align}
Then there are non-singular poles, which appear when two propagators become equal in the $q\rightarrow 0$ limit:
\begin{figure}[H]
\centering
\includegraphics[scale=0.5]{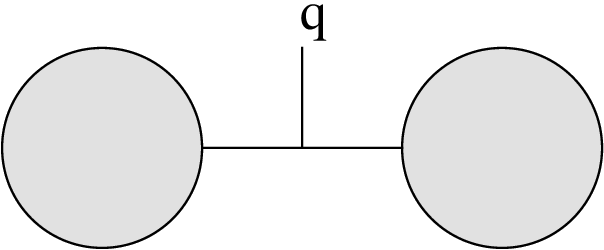}
\end{figure}
\noindent and can be written as:
\begin{align}
D_{n+1}=\frac{N}{P_L(q)(p_1+p_2+\ldots+p_i)^2(q+p_1+p_2+\ldots+p_i)^2 P_R(q)}\, ,
\end{align}
where for Yang-Mills $P_i^2=(p_1+p_2+\ldots+p_i)^2$ contains only consecutive momenta up to particle $i$, $i=\overline{2,n-2}$. We will factor out one of the $P_i^2$'s, and incorporate the other into the lower point local function $B_n$:
\begin{align}\label{d2}
D_{n+1}\rightarrow \frac{N}{(p_1+p_2+\ldots+p_i)^2\mathcal{P}_n(q)}=\frac{1}{(p_1+p_2+\ldots+p_i)^2}B_n\, .
\end{align}
The argument by induction can be used precisely because of this factorization into the lower point local functions.

\subsection{Yang-Mills}

\paragraph{Leading order} The leading $z^{-1}$ piece of the soft limit (\ref{Taylor}) can only come from $q$-pole terms. Using linearity in $e_{n+1}=e$, it can be written as:
\begin{align}
M_{n+1}^{-1}(p^{n-1})= \frac{e^\mu B^\mu_{n}(p^{n-1})}{q.p_1}+\frac{e^\mu C^\mu_{n}(p^{n-1})}{q.p_n}\, ,
\end{align}
where $B_n^\mu$ and $C_n^\mu$ are local (vectors) at $n$-points. Next, gauge invariance in $q$ requires $B^\mu_n=p_1^\mu B_n$, and $C^\mu_n=-p_n^\mu B_n$, where $B_n$ is a local function at $n$ points. Then the leading piece is:
\begin{align}
M_{n+1}^{-1}(p^{n-1})= \left(\frac{e.p_1}{q.p_1}-\frac{e.p_n }{q.p_n}\right)B_n(p^{n-2})\, .
\end{align}
By assumption, gauge invariance in the remaining $(n-1)$ particles uniquely fixes $B_n=A_n$, reproducing the well known Weinberg soft factor \cite{Weinberg}. Note that unlike other methods of obtaining the soft term, we have not used any information on factorization, but only gauge invariance.

Now that the leading order is fixed, consider instead the function $B_{n+1}=M_{n+1}-A_{n+1}$. $B_{n+1}$ is also local, and must be gauge invariant in $n$ particles, but has a vanishing leading order. We will show that all higher orders in the soft expansion of $B_{n+1}$ also vanish, implying that $M_{n+1}=A_{n+1}$. This procedure will be identical for gravity, NLSM and DBI.

\paragraph{Sub-leading order}
Because the leading order vanishes, the sub-leading piece is given only by:
\begin{align}\label{sub1}
B_{n+1}^0(p^{n-1})=\sum_{i=1;n} \frac{e^\mu q^\nu B^{\mu\nu}_{n;i}(p^{n-2}) }{q.p_i}+\sum_{i=2}^{n-2} \frac{e^\mu C_{n;i}^\mu(p^{n-1})}{P_i^2}\, ,
\end{align}
which includes both singular and non-singular pole parts.
The non-singular pole terms are ruled out by gauge invariance in $q$, while the $q$-pole terms must be proportional to $e^{[\mu}q^{\nu]}$. Bringing everything under a common denominator we can write
\begin{align}
B_{n+1}^0(p^{n-1})\propto e^{[\mu}q^{\nu]} \left( q.p_n B^{\mu\nu}_{n;1}(p^{n-2})+q.p_1 B^{\mu\nu}_{n;n}(p^{n-2})\right)\equiv e^{[\mu}q^{\nu]}{B'_n}^{\mu\nu}(p^{n-2})\, ,
\end{align}
where ${B'_n}^{\mu\nu}(p^{n-2})$ is a linear combination of tensors with $k=n-2$, so is ruled out by requiring gauge invariance in $n-1$ particles. Therefore $B_{n+1}^0=0$.

\paragraph{Sub-sub-leading order}
The sub-sub-leading piece is given by:
\begin{align}
B_{n+1}^1(p^{n-1})=e^{[\mu}q^{\nu]}\left( \sum_{i=1,n}\frac{q^\rho B^{\mu\nu\rho}_n(p^{n-3})}{q.p_i}+\sum_{i=2}^{n-2}\frac{C^{\mu\nu}_n(p^{n-2}) }{P_i^2}\right)\, .
\end{align}
This time the non-singular pole terms are not ruled out just by gauge invariance in $q$. We can still write:
\begin{align}
B_{n+1}^1(p^{n-1})\propto e^{[\mu}q^{\nu]} {B'_n}^{\mu\nu}(p^{n-2})\, .
\end{align}
We obtain similar constraints as in the subleading case, which imply $B_{n+1}^1=0$.

\paragraph{Sub$^{s\ge 3}$-leading order}
At arbitrary order $s\ge 3$ we can write:
\begin{align}
B_{n+1}^{s-1}(p^{n-1})\propto e^{[\mu}q^{\nu]}q^{\rho_1}\ldots q^{\rho_{s-2}} B^{\mu\nu\rho_1\ldots \rho_{s-2}}_n(p^{n-s})\, .
\end{align}
And all constraints will have $k\le n-3$, with $\Delta=s-1\ge 1$, so $B^{s-1}_{n+1}=0$ to all orders up to $s=n$, where the soft expansion terminates. Therefore $M_{n+1}=A_{n+1}$, proving uniqueness.

\subsection{Gravity}

For gravity, we can simply write polarization tensors in terms of polarization vectors as $e_i^{\mu\nu}=e_i^\mu f_{i}^\nu$. Then gauge invariance in one graviton becomes equivalent to gauge invariance in two ``gluons". The polynomial statement from section 2 still applies, so ignoring momentum conservation, no polynomial with at most $k$ $e.p$ factors can be gauge invariant in $k+1$ ``gluons''. With momentum conservation, in the case of gravity this is true for $k< 2n-4$. This implies that a tensor $B_n^{\mu\nu}$ with $k$ powers of momenta in the numerator cannot be gauge invariant in $k+1$ particles for $k\le 2n-4$.

One other difference is that for gravity we are no longer restricted only to cyclic poles, since there is no ordering. In the end, the proof is almost identical to that for Yang-Mills. We assume that $A_n(p^{2n-4})$ is unique, and prove the same is true for $A_{n+1}(p^{2n-2})$.
\paragraph{Leading order} The leading piece has a form:
\begin{align}
M_{n+1}^{-1}=\sum_i \frac{e^\mu f^\nu B_{n;i}^{\mu\nu}}{q.p_i}\, .
\end{align}
Gauge invariance in $e$ and $f$ can only be satisfied on the support of momentum conservation, by $B^{\mu\nu}_{n;i}=p_i^\mu p_i^\nu B_n$, where $B_n$ is a local function at $n$ points. Then the leading piece is:
\begin{align}
M_{n+1}^{-1}(p^{2n-2})=\sum_i \frac{e.p_i f.p_i}{q.p_i}B_n(p^{2n-4})\, ,
\end{align}
and now by assumption gauge invariance in the other particles fixes $B_n=A_n$. Using the same trick as before, we consider instead the function $B_{n+1}=M_{n+1}-A_{n+1}$.
\paragraph{Sub-leading order}
The subleading piece is given by:
\begin{align}
B_{n+1}^0=\sum_{i} \frac{e^\mu f^\nu q^\rho B^{\mu\nu\rho}_{n;i}}{q.p_i}+\sum_{i} \frac{e^\mu f^\nu C_{n;i}^{\mu\nu}}{P_i^2}\, .
\end{align}
Gauge invariance in $e$ and $f$ rules out the non-singular pole contributions, and fixes the first term to:
\begin{align}
\nonumber B_{n+1}^0(p^{2n-2})&=\sum_{i} \frac{e^{[\mu} q^{\nu]} f.p_i B^{\mu\nu}_{n;i}(p^{2n-4})}{q.p_i}\\
&=e^{[\mu} q^{\nu]} B^{\mu\nu}_n(p^{2n-4})\, ,
\end{align}
which is ruled out by gauge invariance in the remaining particles.
\paragraph{Sub-sub-leading order}
For higher orders, which go up to $s=2n-1$, the same argument rules out any other solutions, so $A_{n+1}$ is uniquely fixed by gauge invariance.

\section{Unitarity from locality and the Adler zero}\label{unique2}

For the NLSM and DBI we will also deal with local functions $B_n(p^k)$, with $k$ powers of momenta in the numerator. However, the poles are now associated to propagators of quartic diagrams, ordered for the NLSM, un-ordered for DBI. The Adler zero condition \cite{adler} states that the amplitude $A_n$ must vanish when a particle is taken soft. Exactly how rapidly it must vanish sets the difference between the NLSM and DBI \cite{trnka1}-\cite{trnka4}. The limit $p_i\rightarrow 0$ is taken as $p_i=w_i p_i$, $w_i\rightarrow 0$. Then for the NLSM we require the amplitude to vanish as $\mathcal{O}(w_i)$, while for DBI we require $\mathcal{O}(w_i^2)$, $\forall i\neq 3$.

As for gauge invariance, it will be useful to quantify the difference between available momenta and total constraints. In general, if we require a function $B_n(p^k)$ to vanish as $\mathcal{O}(w_i^{g_i})$ for some particle $i$, let the corresponding constraint be $g_i$. Then $g=\sum_i g_i$ will the total constraints $B_n(p^k)$ must satisfy, and define $\Delta=g-k$ as before. This time we wish to show that the NLSM amplitude $A_n(p^{n-2})$ is the unique object satisfying $\Delta=1$ constraints, while the DBI amplitude $A_n(p^{2n-4})$ uniquely satisfies $\Delta=2$ constraints.

We will also show this by counting possible solutions, order by order in the double soft expansion $q= z q$, $\tilde{q}= z \tilde{q}$, $z\rightarrow 0$. The double soft expansion is chosen now, since the functions simply vanish in the single soft limit. The proof will again be almost identical with the ones for YM and GR, with one important difference. In the first two cases, the simple polynomial statement of section 2 significantly streamlined the argument. Remember that in the soft limit, we encountered tensors $B^{\mu\nu}(p^k)$, with $k\le n-2$, which were immediately ruled out. The key in that case was that we could always associate a polynomial with $k$ $e.p$ factors to a function with $k$ momenta in the numerator. For the NLSM, there is no such (simple) distinction to be made, since all we have are $p_i.p_j$ factors, both in the numerator and denominator.

Therefore, for scalars we do not have a direct proof for the following fact: a function $B_n(p^k)$ cannot satisfy $k+1$ constraints, if $k<n-2$. Instead, this statement must be proven by induction. We will write the proof only for the uniqueness statement, ie $k=n-2$, under the assumption that the non-existence statement is true. The proof for the latter case is identical, only with $k<n-2$.

The Taylor series expansion is identical to eq. (\ref{Taylor}). We have the singular $q$-pole terms:
\begin{figure}[H]
\centering
\includegraphics[scale=0.5]{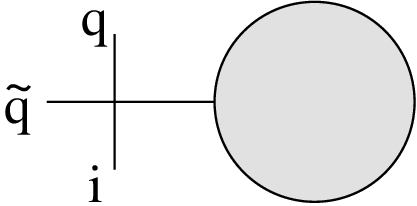}
\end{figure}
\noindent of the form:
\begin{align}
D_{n+2}=\frac{N}{(q+\tilde{q}+p_i)^2 \mathcal{P}_n(q,\tilde{q})}
\end{align}
\noindent where $i=1$ or $i=n$ for NLSM due to ordering. In the soft limit we can also write this in terms of the lower point local function:
\begin{align}
D_{n+2}\rightarrow \frac{1}{(q+\tilde{q}).p_i} \frac{N}{\mathcal{P}_n(0,0)}=\frac{1}{(q+\tilde{q}).p_i}B_n\, .
\end{align}
Next there are non-singular poles, which are more varied than in the cubic diagram case. There is still the equivalent of the double poles from before:
\begin{figure}[H]
\centering
\includegraphics[scale=0.5]{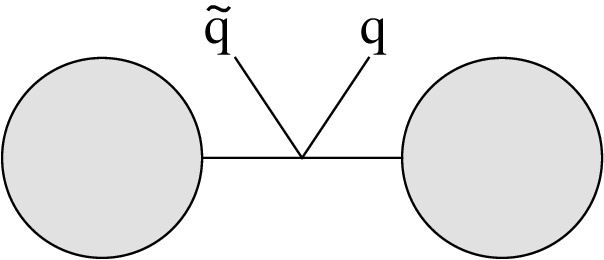}
\end{figure}
\noindent which we can write as:
\begin{align}
D_{n+2}=\frac{N}{P_L(q,\tilde{q})(p_1+p_2+\ldots+p_i)^2(q+\tilde{q}+p_1+p_2+\ldots+p_i)^2 P_R(q,\tilde{q})}\, ,
\end{align}
In the soft limit this becomes:
\begin{align}
D_{n+2}\rightarrow \frac{1}{(p_1+p_2+\ldots+p_i)^2}B_n\, .
\end{align}
There are also more complicated non-singular poles, when the $q$ and $\tilde{q}$ legs are separated. However, even in such cases it is easy to write the terms in a form:
\begin{align}
D_{n+2}\rightarrow \frac{1}{P_i^2}B_n\, .
\end{align}

\subsection{NLSM}

\paragraph{Leading order}
The leading $1/z$ term can only come from $q$-pole terms:
\begin{align}
M_{n+2}^{-1}=\frac{N_1(0,0)}{p_1.(q+\tilde{q})}+\frac{N_2(0,0)}{p_n.(q+\tilde{q})}\, ,
\end{align}
imposing vanishing under $\tilde{q}\rightarrow 0$ implies:
\begin{align}
\left(\frac{N_1}{q.p_1}+\frac{N_2}{q.p_n}\right)=0\, ,
\end{align}
so $N_1=N_2=0$, and $M_{n+2}^{-1}=0$.

\paragraph{Sub-leading order}
At this level both types of poles can contribute. The $q$-pole piece is:
\begin{align}
M_{n+2}^0=\frac{q^\mu B^\mu_{n}+{\tilde{q}}^\mu C^\mu_{n}}{(q+\tilde{q}).p_1}+\frac{q^\mu D^\mu_{n}+{\tilde{q}}^\mu E^\mu_{n}}{(q+\tilde{q}).p_n}\, .
\end{align}
Vanishing under $\tilde{q}\rightarrow 0$ implies $B^\mu=p_1^\mu B_n$, and $D_n^\mu=-p_n^\mu B_n$, and similarly $q\rightarrow 0$ leads to $C_n^\mu=p_1^\mu C_n$ and $E_n^\mu=-p_n^\mu C_n$. The subleading term becomes:
\begin{align}
M^0_{n+2}(p^n)=\left(\frac{q.p_1}{(q+\tilde{q}).p_1}-\frac{q.p_n}{(q+\tilde{q}).p_n}\right)(B_n-C_n)\, .
\end{align}
Now $(B_n-C_n)\equiv B_n(p^{n-2})$ is also a general local function at $n$-points, so by assumption vanishing in the other soft limits fixes $B_n=A_n$.

Terms with non-singular poles have a form:
\begin{align}
\sum_i \frac{N_i}{P_i}\, ,
\end{align}
but are quickly ruled out by requiring vanishing under $q$ or $\tilde{q}$.

\paragraph{Sub-sub-leading order}

The most general $q$-pole sub-sub-leading term is:
\begin{align}
\nonumber M_{n+2}^1=&\frac{1}{(q+\tilde{q}).p_1}(q^\mu q^\nu B^{\mu\nu}+q^\mu \tilde{q}^\nu C^{\mu\nu}+\tilde{q}^\mu \tilde{q}^\nu D^{\mu\nu}+ q.\tilde{q} E)\\
&-\frac{1}{(q+\tilde{q}).p_n}(q^\mu q^\nu F^{\mu\nu}+q^\mu \tilde{q}^\nu G^{\mu\nu}+\tilde{q}^\mu \tilde{q}^\nu H^{\mu\nu}+ q.\tilde{q} I)\, .
\end{align}
Now we expand the remaining $p_i=w_i p_i$ and require $M_{n+2}^1\propto \mathcal{O}(w_i^{1})$ for each of the ($n-1$) particles left. All of the above functions can be treated as independent because of their unique prefactors. Taking the $p$'s in the denominator into account, we obtain the following constraints for all functions:
\begin{align}
B^{\mu\nu},\, C^{\mu\nu}\, D^{\mu\nu},\, E&\propto \mathcal{O}(w_1^2), \mathcal{O}(w_{i\neq 1}^1)\, ,\\
F^{\mu\nu},\, G^{\mu\nu}\, H^{\mu\nu},\, I&\propto \mathcal{O}(w_n^2), \mathcal{O}(w_{i\neq n}^1)\, ,
\end{align}
while component-wise there will be two types of constraints. First:
\begin{align}
\label{nlsm1}
B(p^{n-4}),C(p^{{n-4}}),D(p^{n-4}),G(p^{n-4}) \propto n-2\, .
\end{align}
These are $\Delta=2$ with $k<n-2$ so are ruled out. The other constraints are:
\begin{align}
E(p^{n-2}),\ I(p^{n-2})\propto n\, .
\end{align}
First, we can use $n-1$ of the usual $\mathcal{O}(w_i^1)$ constraints to fix $E=I=A_n$. But then $A_n$ cannot satisfy the extra requirement of $\mathcal{O}(w_1^2)$ or $\mathcal{O}(w_n^2)$, so it must mean that $I=E=0$.

Non-singular poles are still ruled out by vanishing under $q$ and $\tilde{q}$, and so $M_{n+2}^1=0$.

\paragraph{Sub$^{s\ge3}$-leading} With each extra $q^\mu$ or $\tilde{q}^\mu$ being added, $k$ decreases by 1, so $\Delta$ can only increase by at least 1, leading to $\Delta_s=s\ge 2$ constraints. Therefore any sub$^s$-leading order vanishes, and so $M_{n+2}=0$, proving our statement, with the caveat below.

\subparagraph{Neutral poles}
At the sub$^{3}$-leading order some special combinations of non-singular pole terms are not directly ruled out. Consider for example the two diagrams, which we take to have equal numerators:
\begin{figure}[H]
\centering
\includegraphics[scale=0.75]{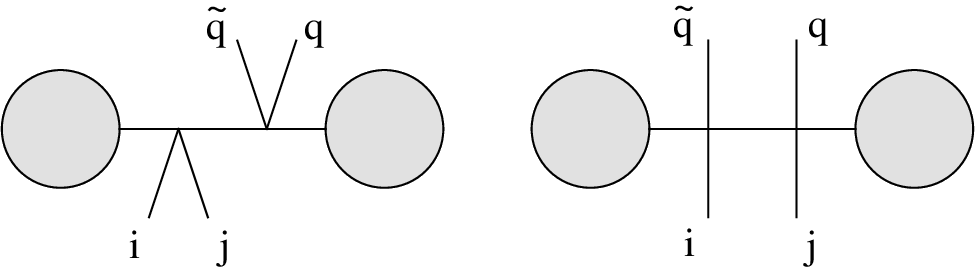}
\end{figure}
given by:
\begin{align}
\nonumber D_{n+2}(p^n)=&\frac{N(p^{n})}{P_L^2 (P_L+i+j)^2 (P_L+i+j+q+\tilde{q})^2P_R^2(q,\tilde{q})}\\
&-\frac{N(p^n)}{P_L^2 (P_L+i+\tilde{q})^2 (P_L+i+j+q+\tilde{q})^2P_R^2(q,\tilde{q})}\, .
\end{align}
At sub$^{s\ge3}$-leading order their contribution is:
\begin{align}\label{extranlsm}
D_{n+2}^{2}(p^n)=q^\mu \tilde{q}^\nu N^{\mu\nu}(p^{n-2})\left(\frac{1}{P_L^2 (P_L+p_i+p_j)^2P_R^2}-\frac{1}{P_L^2 (P_L+p_i)^2P_R^2}\right)\, .
\end{align}
Now $N^{\mu\nu}(p^{n-2})$ has enough momenta to trivially satisfy $n-2$ of the remaining $n-1$ constraints. But vanishing in particle $j$ is automatic because the two denominators in (\ref{extranlsm}) become equal when $p_j\rightarrow 0$. Therefore $D^2_{n+2}$ is not ruled out by our usual argument. Instead, such terms can be ruled out by taking different soft limits. Specifically, it must be soft limits which lead to soft-singularities in $P_L^2$ or $P_R^2$. This ensures that $D^2_{n+2}$ above avoids non-singular pole terms in the new soft limit.

\subsection{Dirac-Born-Infeld}

For DBI we can use the same notation from the previous section. In this case, the Adler zero condition is stronger, as we require $A_n\propto \mathcal{O}(w_i^2)$ under $p_i=w_i p_i\rightarrow 0$. The proof is identical to the one for the NLSM, with the minor difference that now non-cyclic poles are allowed. Also, in all cases non-singular poles can be ruled out easily - the issue appearing in the NLSM is not present, since the vanishing under $p_j\rightarrow 0 $ ensured by eq. (\ref{extranlsm}) can not provide the full $\mathcal{O}(w_i^2)$ needed. Instead, there is a different issue appearing at the same order, which can be resolved by demanding permutation invariance.

\paragraph{Leading order}
The leading piece is given by:
\begin{align}
B^{-1}_{n+2}=\sum_{i=1}^n \frac{N_i}{(q+\tilde{q}).p_i}\, ,
\end{align}
but is ruled out by requiring $B_{n+2}^{-1}\propto \mathcal{O}(w^1)$ in $q$ or $\tilde{q}$\, .

\paragraph{Sub-leading order}
Regular $q$-pole terms have the form:
\begin{align}
\sum_{i=1}^n\frac{1}{(q+\tilde{q}).p_i}(q^\mu B_i^\mu+\tilde{q}^\mu C_i^\mu)\, ,
\end{align}
but are ruled out by requiring $\mathcal{O}(w^2)$ under $q$ and $\tilde{q}$.

\paragraph{Sub-sub-leading order}
Have the form:
\begin{align}
M^1_{n+2}&=\sum_i\frac{1}{(q+\tilde{q}).p_i}(q^\mu q^\nu\, B_i^{\mu\nu}+\tilde{q}^\mu \tilde{q}^\nu\, C_i^{\mu\nu}+q^\mu \tilde{q}^\nu D_i^{\mu\nu}+q.\tilde{q} E_i )\, .
\end{align}
Requiring $\mathcal{O}(w^2)$ in $q, \tilde{q}$ we end up with:
\begin{align}
\nonumber M^1_{n+2}(p^{2n})&=\sum_i\frac{1}{(q+\tilde{q}).p_i}((q.p_i)^2 B+(\tilde{q}.p_i)^2 C+q.p_i \tilde{q}.p_i D )\\
&=\sum_i\frac{q.p_i \tilde{q}.p_i}{(q+\tilde{q}).p_i} (-B-C+ D )\, .
\end{align}
Now $(-B-C+D)\equiv B_n(p^{2n-4})$ is a general local function at $n$-points, so imposing the remaining $2n-2$ constraints fixes $B_n=A_n$ by assumption.

\paragraph{Sub$^3$-leading order}
Like for the NLSM, at this order extra care is required. The usual arguments rule out all terms except:
\begin{align}
D_{n+2}^2(p^{2n})= q.\tilde{q}\sum_i \frac{ q.p_i B_{n;i}+\tilde{q}.p_i C_{n;i}}{(q+\tilde{q}).p_i}\, ,
\end{align}
under the condition that $\sum_i B_{n;i}=\sum_i C_{n;i}=0$. The functions $B_i(p^{2n-4})$ and $C_i(p^{2n-4})$ must satisfy $\Delta=2$ constraints and are fixed (up to some coefficient) to $A_n$ by assumption. Then the extra conditions become $\sum_i B_{n;i}=\sum_i b_i A_n=0$, and similarly $\sum C_i=\sum_i c_i A_n=0$. The sub$^3$-leading term becomes:
\begin{align}
D_{n+2}^2= q.\tilde{q} A_n \sum_i\frac{b_i \,q.p_i+c_i\, \tilde{q}.p_i}{(q+\tilde{q}).p_i}\, .
\end{align}
But now if we require symmetry in $q\leftrightarrow \tilde{q}$, then $b_i=c_i$, so $D_{n+2}^2=0$, and this order vanishes.
\paragraph{Sub$^{s>3}$leading}
All such terms are ruled out, so $M_{n+2}=0$ to all orders, and $A_{n+2}$ is unique.

\section{Locality and unitarity from singularities and gauge invariance}\label{locality}

The general argument we used in the previous sections can be easily extended when we relax our cubic graph assumptions, and instead consider a more general singularity structure, as long as the singularities themselves have a form $\left(\sum_i p_i\right)^2$, with consecutive momenta in the case of Yang-Mills. This means we allow double poles as well as overlaps. There are three cases to consider depending on how many singularities $s$ we allow. We will show that:
\begin{itemize}
\item for $s<n-3$ there is no solution
\item for $s=n-3$ there is a unique solution, $A_n$
\item for $s> n-3$ solutions can be factorized in a form $\left(\sum \textrm{poles} \right)\times A_n$
\end{itemize}
We prove these three results for five points Yang-Mills. It is easy to extend the proof for general $n$, including for gravity.

In the following, we will call a function with $s$ singularities $B_{n;s}$, and to simplify notation, let:
\begin{align}
S_0=\frac{e.p_1}{q.p_4}-\frac{e.p_4}{q.p_4}\, .
\end{align}

\subsection{Case 1. $s<n-3$}
This case is easy to prove by induction. Assume that $B_{4;0}(p^{2})$ is ruled out by gauge invariance. Then the five point function with just one singularity has a leading order:
\begin{align}
M_{5;1}^{-1}(p^3)=S_0 B_{4;0}(p^2)\, ,
\end{align}
which by assumption is ruled out. Higher order terms are again ruled out as usual, so there are no solutions for $s< n-3$.

\subsection{Case 2. $s=n-3$}
At five points, in this case we have two (cyclic) poles per term, and now we also allow double poles and overlaps.
\paragraph{Order $z^{-2}$}
The lowest order is now $z^{-2}$, coming from three possible terms, which were not present before:
\begin{align}
M_{5;2}^{-2}(p^3)=\frac{N_a}{(q.p_1)^2}+\frac{N_b}{(q.p_1)(q.p_4)}+\frac{N_c}{(q.p_4)^2}\, .
\end{align}
Gauge invariance in $q$ requires the forms:
\begin{align}
M_{5;2}^{-2}(p^3)=\frac{1}{q.p_1}S_0 B_{4;0}(p^2)+\frac{1}{q.p_4}S_0 C_{4;0}(p^2)\, ,
\end{align}
so both $B_{4;0}$ and $C_{4;0}$ vanish by the previous argument.

\paragraph{Order $z^{-1}$} At this order we have the usual leading piece, but also terms with the non-local poles from above:
\begin{align}
\nonumber M_{5}^{-1}=& S_0 B_4(p^{2})+e^{[\mu}q^{\nu]} \left(\frac{N_a^{\mu\nu}}{(q.p_1)^2}+\frac{N_b^{\mu\nu}}{(q.p_1)(q.p_4)}+\frac{N_c^{\mu\nu}}{(q.p_4)^2}\right)\, .
\end{align}
For the second piece we need tensors $N_4^{\mu\nu}(p^2)$, gauge invariant in three particles, which is not possible. Therefore the leading piece is just
\begin{align}\label{un}
M_{5}^{-1}=S_0A_4\, ,
\end{align}
and so far we get the same answer as usual. However, we must deal with a subtle issue that was not present before. We have shown that at the leading order, all possible functions must map onto the unique expression (\ref{un}). But when we allow a non-local singularity structure, it is possible for two different $n+1$ point functions to have an identical leading order piece. Consider for example the actual amplitude, which contains a local term such as:
\begin{align}\label{good}
A_5=\ldots \frac{e.p_1 N}{q.p_1 (q+p_1+p_2)^2}+\ldots \, ,
\end{align}
and a similar function $M_5$, but where we replace the term from above with a non-local one:
\begin{align}\label{bad}
M_5=\ldots \frac{e.p_1 N}{q.p_1 (p_1+p_2)^2}+\ldots\, .
\end{align}
In the soft limit $q\rightarrow 0$ both functions are equal at the leading order, so apparently we have two different solutions, contradicting our statement. The issue can still be resolved by considering all orders of the soft expansion of $B_5=M_5-A_5$. The subleading order is now different than the usual (\ref{sub1}), because $B_5$ now has a contribution originating from the Taylor series expansion of the denominator in eq. (\ref{good}), which was absent before. We obtain:
\begin{align}
B_5^{0}=\frac{e^\mu q^\nu B^{\mu\nu}_4}{q.p_i}+e^\mu C_4^\mu -\frac{e.p_iq.(p_1+p_2) N}{q.p_i(p_1.p_2)^2}+\ldots
\end{align}
where the third term is new. But using our previous arguments $B_5^{0}$ is still ruled out by gauge invariance. Higher order terms can be treated in a similar manner, so $B_5=0$ to all orders. Therefore the five point Yang-Mills amplitude is completely fixed even if we start with these non-local assumptions.

\subsection{Case 3. $s> n-3$}
For this case, where we are not expecting to obtain a unique answer, but the same soft limit argument can be used to count the maximum total number of independent solutions, order by order. First, at four points it is easy to check that with $s=2$ poles, there are two solutions:
\begin{align}
M_{4;2}=\left(\frac{a_1}{p_1.p_2}+\frac{a_2}{p_1.p_4}\right)A_4\, .
\end{align}
Now at five points, with three poles, we want to show there are five solutions, corresponding to the five different cyclic poles:
\begin{align}\label{five}
M_{5;3}=\left(\frac{a_1}{p_1.p_2}+\frac{a_2}{p_2.p_3}+\ldots+\frac{a_5}{p_5.p_1}\right)A_5\, .
\end{align}

Again taking a soft limit, and imposing gauge invariance in $p_5=q$, we obtain:
\paragraph{Order $\mathcal{O}(z^{-3})$}
\begin{align}
M_{5;3}^{-3}= \frac{1}{(q.p_1)^2}S_0 B_{4;0}+\frac{1}{(q.p_4)^2}S_0 C_{4;0}\, ,
\end{align}
which was shown to vanish, so no solutions at this level.
\paragraph{Order $\mathcal{O}(z^{-2})$}
\begin{align}
M_{5;3}^{-2}=\frac{1}{q.p_1}S_0 B_{4;1}+\frac{1}{q.p_4}S_0 C_{4;1}\, ,
\end{align}
which is fixed by gauge invariance to
\begin{align}
M_{5;3}^{-2}=\left(\frac{a_5}{q.p_1}+\frac{a_4}{q.p_4}\right)S_0 A_4\, .
\end{align}
Therefore from this order we obtain two possible solutions.
\paragraph{Order $\mathcal{O}(z^{-1})$}
Because we are only counting {\it independent} solutions, we can simply ignore the contributions from the lower order above. Therefore we are only interested in the term:
\begin{align}
M_{5;3}^{-2}=S_0 B_{4;2}\, .
\end{align}
By assumption this gives two independent solutions corresponding to the poles $p_1.p_2$ and $p_1.p_4$, but starting from five points there are three poles which map onto these two in the soft limit:
\begin{align}
&p_1.p_2\rightarrow p_1.p_2\, ,\\
&p_2.p_3=(p_4+p_5+p_1)^2\rightarrow p_1.p_4\, ,\\
&p_3.p_4=(p_5+p_1+p_2)^2\rightarrow p_1.p_2\, .
\end{align}
And so we obtain three independent solutions at this order. For higher orders, the usual arguments rule out other solutions, and so we end up with at most five possible solutions. We have not derived what these must be, but since we can just write down the five terms of Eq. (\ref{five}), this must be all of them. The result is easy to generalize to an arbitrary number of extra poles.

The argument can also easily be extended to general $n$-point amplitudes, as well as gravity. Once it is shown that functions with $s<n-3$ singularities are ruled out, for $s=n-3$ the only non-vanishing contribution will be the Weinberg term at order $1/z$, which by the usual argument implies uniqueness. We suspect the same type of argument can be used for NLSM and DBI, although some extra complications might appear at the sub$^3$-leading orders, which were already troublesome. Regardless, a more direct argument ruling out the non-local terms was already presented in Ref. \cite{Nima} for these theories.


\section{Generalizing singularities}\label{l2}
\subsection{Non-local singularities}
In the previous sections we have assumed that the denominators are always products of singularities $P_\mathcal{S}^2=\left(\sum_i p_i\right)^2$. An obvious next step is to relax even this assumption, and allow completely non-local singularities of the form $(\sum_i a_i p_i)^2$. In full generality, this doesn't work out. Even at four points, allowing a singularity of the form $a\, p_1.p_2+b\, p_1.p_4$ no longer provides a unique local solution.  We can write the four point numerator as $N_4=  (t N_s+ s N_t)=(t,s)\cdot(N_s,N_t)$, with $A_4=N_4/(st)$. Now we can do any 2D rotation to obtain $N_4=(t',s')\cdot(N'_s,N'_t)$, where $s'=s\, \textrm{cos}\, \theta - t\, \textrm{sin}\,\theta $ and $t'=t\, \textrm{cos}\, \theta+s\, \textrm{sin}\,\theta$. But now diving by $(s't')$ we obtain a (non-unique) gauge invariant with the non-local poles $s'$ and $t'$, so our claim is invalidated if we allow such poles.

However, there exists a special set of non-local ``cyclic'' poles from which full locality can still be derived, if we are careful about momentum conservation. To obtain this set, we must start from a local cyclic pole $P_{jk}^2=(\sum_{i=j}^k p_i)^2$. Now only {\it after} using momentum conservation  $p_3=-\sum p_i$, we can add arbitrary coefficients $(\sum_i p_i)^2\rightarrow (\sum_i a_i p_i)^2$. For example, from a six point local pole like $(p_1+p_2+p_3)^2$, we can obtain $(a\, p_4+b\, p_5+c\, p_6)^2$. Note how this rule doesn't allow the four point  pole $a\, p_1.p_2+b\, p_1.p_4$ from above. It can only come from the pole $p_1.p_3=p_1.p_2+p_1.p_4$, which is not cyclic. At five points, the most general set of singularities that can be used is:
\begin{align}
\nonumber (p_1+p_2)^2&=p_1.p_2\\
\nonumber(p_2+p_3)^2&=(p_1+p_4+p_5)^2\rightarrow a_1\, p_1.p_4+a_2\, p_1.p_5+a_3\, p_4.p_5\\
\nonumber(p_3+p_4)^2&=(p_1+p_2+p_5)^2\rightarrow a_4\, p_1.p_2+a_5\, p_1.p_5+a_6 \,p_2.p_5\\
\nonumber(p_4+p_5)^2&=p_4.p_5\\
(p_5+p_1)^2&=p_1.p_5
\end{align}
For an $n$ point amplitude, $n-2$ of the singularities keep the form $p_i.p_{i+1}$, while the others are promoted to carry these extra coefficients. Now, the usual proof by induction will work, as long as we avoid taking soft the particles adjacent to 3, which is of course always possible from four points and higher. This procedure ensures that the soft-singularities $q.p_i$, critical for the leading term, are not affected in any way. Then the leading term is as usual
\begin{align}
B_{n+1}^\textrm{non-local}\rightarrow \left(\frac{e.p_1}{q.p_1}-\frac{e.p_n}{q.p_n}\right)B_n^{\textrm{non-local}}\, ,
\end{align}
where now $B_n^{\textrm{non-local}}$ also contains the non-local singularities described above. If by assumption even this non-local $B_n$ is uniquely fixed by gauge invariance, ultimately so will $B_{n+1}$. The claim is in fact trivial at four points, where none of the poles may be modified, so $B_4^{\textrm{non-local}}=B_4^{\textrm{local}}$.

With a few extra restrictions, a similar result can be shown for gravity as well, though the procedure is somewhat more complicated because for gravity soft-singularities involving $p_3$ are not so easily avoided. The solution is to require several extra poles to keep their usual local form, in such a way to ensure that even after taking multiple soft limits, there always exists a particle which forms no soft-singularities with $p_3$.

\subsection{No singularities}\label{nonlocality}

So far, we have mostly looked at functions with singularities of the form $(\sum_i p_i)^2$, and in some cases we showed that singularities of the type $(\sum_i a_i p_i)^2$ also lead to uniqueness. But what about allowing the denominators to be  polynomials of some degree $s^2$, instead of $s$ products of singularities? In general, this is a very difficult question to systematically analyze, and given the four point counter-example from the previous section, it might simply be an ill-posed question. But instead of trying to understand all such completely general poles, there is an even more general alternative to pursue. We can completely disregard singularities, and investigate gauge invariance directly at the level of the total numerator, by considering general polynomials instead of functions with poles. Clearly, given sufficient mass dimension, a general polynomial can always be thought of as originating from the most general singularity structure possible. We can start with the minimal polynomial which admits any solution, which has $n-2$ $e.p$ factors, and $(n-2)^2$ total mass dimension, the same as an actual amplitude numerator. It turns out that imposing our usual $n-1$ gauge invariance constraints does not provide a meaningful solution, but imposing the full $n$ constraints does: we obtain a linear combination of amplitude numerators! The $n^{\rm th}$ extra constraint essentially is required to replace the information we lost by not considering denominators which are products of singularities. From this perspective, the singularities do no play any crucial or physical role, but only provide a useful method of organizing terms in the polynomial. While we do not have a proof for this fact for $n>4$, it is easily testable at five points.  There we obtain six solutions, which are linear combinations of five point amplitude numerators, corresponding to different orderings. Below we provide leading order evidence for this fact.

We can again use our usual soft argument to count the solutions at leading order. First, it is easy to check that imposing all four gauge invariance conditions on the four point polynomial $N_4((e.p)^2,p^4)$ gives a unique solution. This corresponds to the fact that all four point amplitudes have the same numerator. That is, any amplitude can be obtained by dividing the same numerator by the desired propagator structure:
\begin{align}
A(1,2,3,4)=\frac{N}{p_1.p_2 \,p_1.p_4}\, ,\\
A(1,3,2,4)=\frac{N}{p_1.p_3 \,p_1.p_4}\, .
\end{align}
At five points, the leading piece of the general polynomial must have a form:
\begin{align}
N_5((e.p)^3,p^9)=e^{[\mu} q^{\nu]}N^{\mu\nu}(p^8)\, .
\end{align}
After imposing the other four constraints all possible components are ruled out, except the following:
\begin{align}
N_5((e.p)^3,p^9)=S_{12}N_{a}+S_{14}N_{b}+S_{24}N_{c}\, ,
\end{align}
where $S_{ij}=e.p_i q.p_j-e.p_jq.p_i$. Now the $N_i((e.p)^2,p^6)$ must also satisfy the four constraints. First, we can rewrite $N((e.p)^2,p^6)=N((e.p)^2,p^4)\sum_{i,j} a_{ij} p_i.p_j$, after a reshuffling of the coefficients. Then, the constraints imposed on $N((e.p)^2,p^6)$ instead act on $N((e.p)^2,p^4)$, which by assumption is fixed uniquely to the four point numerator. Finally, there are two independent $p_i.p_j$ factors at four points. Therefore we obtain
\begin{align}
N((e.p)^3,p^9)=(a_1 p_1.p_2+a_2 p_1.p_4 )(b_1S_{12}+b_2S_{14}+b_3S_{24}) N_4\, ,
\end{align}
ie. six independent solutions, which are related to the leading pieces of amplitude numerators. Unfortunately, the subleading order is not ruled out so quickly. The $N$ still have enough momenta to provide gauge invariant contributions even at this order:
\begin{align}
N((e.p)^3,p^9)=(a_1.q.p_1+a_2 q.p_2 +a_3 q.p_4)(b_1S_{12}+b_2S_{14}+b_3S_{24}) N_4\, ,
\end{align}
and so the usual argument fails in its simplest form. However, considering all orders, eventually these extra solutions become tied to the original six, and in the end just six solutions are left. The argument becomes even less well suited for higher points, so clearly a better strategy is required.

\section{Summary of the results and future directions}\label{ending}

In this note we have presented the full proofs for some of the uniqueness claims originally made in \cite{Nima}. We summarize these results below. Let $s$ be the number of poles of the form $\left(\sum_i p_i\right)^2$, and $k$ the mass dimension of the numerators.

\paragraph{ Yang-Mills and General Relativity:}
\begin{itemize}
\item Unique solution for $s=n-3$, with $k_{\textrm{YM}}=n-2$, $k_{\textrm{GR}}=2n-4$
\item No solutions for $s$ or $k$ smaller than above
\item Factorized solutions $(\sum \textrm{poles})\times A_n$ for $s$ larger than above
\end{itemize}
\paragraph{NLSM and DBI:}
\begin{itemize}
\item Uniqueness assuming quartic diagrams, with $k_{\textrm{NLSM}}=n-2$, $k_{\textrm{DBI}}=2n-4$
\item No solutions for $k$ smaller than above
\end{itemize}

For Yang-Mills, we also proved that uniqueness holds when allowing specific types of non-local singularities $\left(\sum_i a_i p_i\right)^2$. Finally, we conjectured that general polynomials of minimal mass dimension lead to linear combinations of amplitude numerators, and so to both locality and unitarity.

The next step is understanding how to approach such polynomials with absolutely no singularity structure. It would be very interesting to see if the soft limit argument can be extended even further, or if an even more powerful argument is required. Meanwhile, for NLSM and DBI, it is not even clear what the equivalent claim should be, if it exists. For YM, the number of $e.p$ factors always helped distinguish what $p$'s come from numerators and which come from propagators. For scalar theories, there is no distinction to be made: all the $p$'s are equal. We should note that an equivalent claim for gravity does not exist. It is trivial to obtain many different solutions by gluing together Yang-Mills amplitudes, while there is a unique gravity numerator. Nevertheless, even if the numerator statement is less fundamental than the other results, it is a very useful exercise. After all, thinking about polynomials lead to the crucial results of section \ref{gaugeinv}, so perhaps there is more to be learned from this perspective.

A more important issue to be understood is that of the gram determinant relations. When working in some fixed dimension $D$, at most $D-1$ vectors can be linearly independent ($-1$ because of momentum conservation). For example, if we restrict to 4D, starting at six points, we can express $p_6$ in terms of the other four independent momenta:
\begin{align}\label{lin}
p_6=a\, p_1+b\, p_2+ c\, p_4+d\, p_5
\end{align}
This could allow for different solutions to our requirements. The linear dependence (\ref{lin}) can be viewed as another form of momentum conservation:
\begin{align}\label{mln}
p_3=-(p_1+ p_2+ p_4+ p_5+p_6)
\end{align}
We already saw that adding momentum conservation limited the applicability of our initial polynomial argument to $k< n-2$: at $k=n-2$ momentum conservation allowed for some ``free'' gauge invariants to be formed. Luckily, this was still sufficiently constraining for our purposes. It is not inconceivable, though would be very surprising, that the gram determinant relations could allow such free gauge invariants starting at $k=n-3$ for example.

Ultimately, these results strongly suggest that scattering amplitudes might have a different definition, perhaps geometric, in line with the amplituhedron program \cite{amplituhedron}. A formulation where both this minimal singularity structure and gauge invariance/vanishing in the soft limit are manifest could potentially uncover yet more unknown features of these theories.

\section*{Acknowledgments}
The author would like to thank Nima Arkani-Hamed and Jaroslav Trnka for the insights and many discussions which lead to this work, and Song He for valuable discussions.


\begin{thebibliography}{99}
\bibitem{Nima}
N.~Arkani-Hamed, L.~Rodina and J.~Trnka,
``Locality and Unitarity from Singularities and Gauge Invariance,''
arXiv:1612.02797 [hep-th].
\bibitem{bcfwl}
L.~Rodina,
``Uniqueness from locality and BCFW shifts,''
arXiv:1612.03885 [hep-th].

\bibitem{BCFW} R. Britto, F. Cachazo, and Bo Feng, Nucl. Phys. B {\bf 715}, 499 (2005) [arXiv:hep-th/0412308].
R. Britto, F. Cachazo, B. Feng, and E. Witten, Phys. Rev. Lett. {\bf 94},181602 (2005) [arXiv:hep-th/0501052].
\bibitem{bcj}
Z.~Bern, J.~J.~M.~Carrasco and H.~Johansson,
``New Relations for Gauge-Theory Amplitudes,''
Phys.\ Rev.\ D {\bf 78}, 085011 (2008)
doi:10.1103/PhysRevD.78.085011
[arXiv:0805.3993 [hep-ph]].
\bibitem{CHY1} F. Cachazo, S. He and E. Y. Yuan, “Scattering of Massless Particles in Arbitrary
Dimensions,” Phys. Rev. Lett. 113, no. 17, 171601 (2014) [arXiv:1307.2199 [hep-th]].
\bibitem{CHY2} F. Cachazo, S. He and E. Y. Yuan, “Scattering of Massless Particles: Scalars, Gluons
and Gravitons,” JHEP 1407, 033 (2014) [arXiv:1309.0885 [hep-th]]
\bibitem{CHY3} F. Cachazo, S. He and E. Y. Yuan, “Scattering Equations and Matrices: From Einstein
To Yang-Mills, DBI and NLSM,” JHEP 1507, 149 (2015) [arXiv:1412.3479 [hep-th]]





\bibitem{Weinberg}
S.~Weinberg,
``Infrared photons and gravitons,''
Phys.\ Rev.\ {\bf 140}, B516 (1965).
\bibitem{doublesoftstuff}
F.~Cachazo, S.~He and E.~Y.~Yuan,
``New Double Soft Emission Theorems,''
Phys.\ Rev.\ D {\bf 92}, no. 6, 065030 (2015)
doi:10.1103/PhysRevD.92.065030
[arXiv:1503.04816 [hep-th]].


\bibitem{strom1} A. Strominger, “On BMS Invariance of Gravitational Scattering,” arXiv:1312.2229 [hepth].
\bibitem{strom2} T. He, V. Lysov, P. Mitra and A. Strominger, “BMS supertranslations and Weinberg’s
soft graviton theorem,” arXiv:1401.7026 [hep-th].
\bibitem{BH}
S.~W.~Hawking, M.~J.~Perry and A.~Strominger,
``Soft Hair on Black Holes,''
Phys.\ Rev.\ Lett.\ {\bf 116}, no. 23, 231301 (2016)
doi:10.1103/PhysRevLett.116.231301
[arXiv:1601.00921 [hep-th]].


\bibitem{congkao}H.~Luo and C.~Wen,``Recursion relations from soft theorems,''JHEP {\bf 1603}, 088 (2016)[arXiv:1512.06801 [hep-th]].




\bibitem{boel1} 
  L.~A.~Barreiro and R.~Medina,
  ``RNS derivation of N-point disk amplitudes from the revisited S-matrix approach,''
  Nucl.\ Phys.\ B {\bf 886}, 870 (2014)
  doi:10.1016/j.nuclphysb.2014.07.015
  [arXiv:1310.5942 [hep-th]].

\bibitem{boel2} 
  R.~H.~Boels and R.~Medina,
  ``Graviton and gluon scattering from first principles,''
  arXiv:1607.08246 [hep-th].

\bibitem{strom3}
F.~Cachazo and A.~Strominger,
``Evidence for a New Soft Graviton Theorem,''
arXiv:1404.4091 [hep-th].
\bibitem{delta}
J.~Broedel, M.~de Leeuw, J.~Plefka and M.~Rosso,
``Constraining subleading soft gluon and graviton theorems,''
Phys.\ Rev.\ D {\bf 90}, no. 6, 065024 (2014)
[arXiv:1406.6574 [hep-th]].






























\bibitem{adler}
S.~L.~Adler,
``Consistency conditions on the strong interactions implied by a partially conserved axial vector current,''
Phys.\ Rev.\ {\bf 137}, B1022 (1965).
doi:10.1103/PhysRev.137.B1022






\bibitem{trnka1}
K.~Kampf, J.~Novotny and J.~Trnka,
``Tree-level Amplitudes in the Nonlinear Sigma Model,''
JHEP {\bf 1305}, 032 (2013)
doi:10.1007/JHEP05(2013)032
[arXiv:1304.3048 [hep-th]].


\bibitem{trnka2}
C.~Cheung, K.~Kampf, J.~Novotny and J.~Trnka,
``Effective Field Theories from Soft Limits of Scattering Amplitudes,''
Phys.\ Rev.\ Lett.\ {\bf 114}, no. 22, 221602 (2015)
doi:10.1103/PhysRevLett.114.221602
[arXiv:1412.4095 [hep-th]].



\bibitem{trnka3}
C.~Cheung, K.~Kampf, J.~Novotny, C.~H.~Shen and J.~Trnka,
``On-Shell Recursion Relations for Effective Field Theories,''
Phys.\ Rev.\ Lett.\ {\bf 116}, no. 4, 041601 (2016)
doi:10.1103/PhysRevLett.116.041601
[arXiv:1509.03309 [hep-th]].


\bibitem{trnka4}
C.~Cheung, K.~Kampf, J.~Novotny, C.~H.~Shen and J.~Trnka,
``A Periodic Table of Effective Field Theories,''
arXiv:1611.03137 [hep-th].






\bibitem{amplituhedron}
N.~Arkani-Hamed and J.~Trnka,
``The Amplituhedron,''
JHEP {\bf 1410}, 030 (2014)
[arXiv:1312.2007 [hep-th]].


\end{thebibliography}
\end{document}